\documentclass[supp]{new_tlp} 
\usepackage{amsfonts, amsmath, amssymb, mathrsfs}
\usepackage{mathtools}
\usepackage{graphicx}
\usepackage{epsfig}
\usepackage{epstopdf}
\usepackage{url}
\usepackage{multirow}
\usepackage{supp}

\newcommand{\argmax}{\operatornamewithlimits{argmax}}

\long\def\comment#1{}

 \title[(C)LP for DCOP]
        {Logic and Constraint Logic Programming \\
        for Distributed Constraint Optimization}

  \author[Tiep Le et al.]
         {Tiep Le, Enrico Pontelli, Tran Cao Son, William Yeoh\\
         Department of Computer Science, New Mexico State University\\
         \email{\{tile,epontell,tson,wyeoh\}@cs.nmsu.edu}}

\jdate{March 2013}
\pubyear{2013}
\pagerange{\pageref{firstpage}--\pageref{lastpage}}
\doi{S1471068401001193}

\pubauthor{Tiep Le, Enrico Pontelli, Tran Cao Son, William Yeoh}
\jurl{xxxxxx} 
\pubdate{22 June 2013}

\def\clpdcop{{LP-DPOP}}
\def\clpdcopr{{\small {LP-DPOP$^{\textnormal{rules}}$}}}
\def\clpdcopf{{\small {LP-DPOP$^{\textnormal{facts}}$}}}

\begin{document}

\maketitle

\label{firstpage}

 \begin{abstract}
   The field of \emph{Distributed Constraint Optimization Problems (DCOPs)}
   has gained momentum, thanks
   to its suitability in capturing complex problems (e.g., 
   multi-agent coordination and resource allocation problems) that are naturally distributed and cannot
   be realistically addressed in a centralized manner. The state of the art in 
   solving DCOPs relies on the use of ad-hoc infrastructures and ad-hoc constraint
   solving procedures. This paper investigates an infrastructure for solving DCOPs
   that is completely built on logic programming technologies. In particular, the
   paper explores the use of a general constraint solver (a 
   \emph{constraint logic programming system} in this context) to
   handle the agent-level constraint solving. The preliminary experiments show
   that logic programming provides benefits over a state-of-the-art DCOP system,
   in terms of performance and scalability, opening the doors to the use of more advanced
   technology (e.g., search strategies and complex constraints) for solving DCOPs.
  \end{abstract}


\section{Introduction}
\emph{Distributed Constraint Optimization Problems (DCOPs)} are descriptions of constraint
optimization problems where variables and constraints are distributed among a group of agents, and
where each agent can only interact  with agents that share a common constraint 
\cite{modi:05,petcu:05,yeoh:12}. Researchers
have realized the importance of DCOPs, as they naturally capture  real-world scenarios,
where a collective tries to achieve optimal decisions, but without the ability to collect all information
about resources and limitations into a central solver. For example, DCOPs have been successfully 
used to model domains like resource management 
and scheduling \cite{maheswaran:04a,farinelli:08,leaute:11}, 
sensor networks \cite{fitzpatrick:03,jain:09,zhang:05,zivan:09,stranders:09b}, and smart grids \cite{kumar:09,gupta:13}. 

The DCOP field has grown at a fast pace in recent years.
Several popular implementations of DCOP solvers have been created \cite{leaute:09,sultanik:07,ezzahir:07}. The majority of the existing
DCOP algorithms can be placed in one of three classes. \emph{Search-based} algorithms perform
a distributed search over the space of solutions to determine the optimum \cite{modi:05,gershman:09,zhang:05,DBLP:journals/jair/YeohFK10}. \emph{Inference-based} algorithms, on the other hand, make use of techniques from dynamic programming to propagate
aggregate information among agents \cite{petcu:05,farinelli:08,conf/atal/VinyalsRC09}; these two classes provide a different balance between memory
requirements and number of messages  exchanged. Another class of methods  includes approximated algorithms that rely on \emph{sampling} \cite{ottens:12,nguyen:13}
 applied to the overall search space.

The driving objective of the investigation discussed in this paper is
to understand the role that logic programming can play in solving DCOPs. In particular,
existing popular  DCOP solvers (e.g., the frequently used
FRODO platform \cite{leaute:09}) are ad-hoc systems, with a relatively closed structure,
and making use of ad-hoc dedicated solvers for constraint handling within each agent. 
Thus, a question we intend to address with this paper is whether the use of a general
infrastructure for constraint solving 
within each agent of a DCOP would bring benefits compared to the ad-hoc solutions of the 
existing implementations. We propose a general infrastructure (based on distributed dynamic programming)
for the communication among agents, guaranteeing completeness of the system. 
The platform enables the use of a generic logic programming solver (e.g., a Constraint Logic Programming
system) to handle the local constraints within each agent; the generality of the platform
will also allow the use of distinct logic programming paradigms within each agent (e.g., Answer
Set Programming).

The paper discusses the  overall logic programming  infrastructure, along with the 
details of the modeling of each agent using constraint logic programming. We provide some preliminary
experimental results, validating the viability and effectiveness of this research direction for 
DCOPs. The results also highlight the potential offered by logic programming to provide an implicit
representation of hard constraints in DCOPs, enabling a more effective pruning of the 
search space and reducing  memory requirements.
\vspace{-0.3cm}
\section{Background}
In this section, we provide a brief review of basic concepts from DCOPs. We assume that the readers
have familiarity with logic and constraint logic programming; in particular, we will refer 
to the syntax of the {\sf clpfd} library of SICStus Prolog \cite{sicstus}.

\vspace{-0.3cm}
\subsection{Distributed Constraint Optimization Problems (DCOPs)}

A \emph{DCOP}~\cite{modi:05,petcu:05,yeoh:12} is described 
by a tuple ${\cal P}=(X, D, F, A, \alpha)$ where:
{\bf (i)} $X=\{x_1,\dots,x_n\}$ is a set of \emph{variables}; 
{\bf (ii)} ${D} =\{D_{x_1},\ldots,D_{x_n}\}$ is a set of finite \emph{domains},
 where each $D_{x_i}$ is the domain of variable $x_i$; 
{\bf (iii)} ${F} = \{f_1, \ldots, f_m\}$ is a set of \emph{utility functions} (a.k.a. \emph{constraints}), 
where each $f_j: D_{x_{j1}} \times D_{x_{j2}} \times \ldots \times D_{x_{jk}} \mapsto \mathbb{N} \cup \{-\infty, 0\}$ 
specifies the utility of each combination of values of variables in its \emph{scope}
$scp(f_j) = \{x_{j1},\dots, x_{jk}\} \subseteq X$; 
{\bf (iv)} ${A} = \{a_1, \ldots, a_p\}$ is a set of \emph{agents}; and 
{\bf (v)} $\alpha: {X} \rightarrow {A}$ maps each variable to an agent. 

We assume the domains $D_x$ to be finite intervals of integer numbers.
A \emph{substitution} $\theta$
of a DCOP $\cal P$ is a value assignment for  the variables in $X$ s.t. $\theta(x)\in D_x$ for
each $x\in X$.
Its utility is $ut_{\cal P}(\theta)=\sum_{i=1}^m f_i(scp(f_i)\theta)$, i.e., the evaluation of all utility functions on it. A solution $\theta$ is a substitution such that
$ut_{\cal P}(\theta)$ is maximal, i.e., there is no other substitution $\sigma$ such that
$ut_{\cal P}(\theta) < ut_{\cal P}(\sigma)$.
 $Soln_{\cal P}$ denotes the set of solutions of  $\cal P$. 

Each DCOP $\cal P$ is associated with a  \emph{constraint graph}, denoted with $G_{\cal P} = (X, E_{\cal P})$,
where $X$ is a set of nodes  which correspond to DCOP
variables, and $E_{\cal P}$ is a set of edges which connect pairs of variables in the scope of the same utility function.

\vspace{-0.3cm}
\subsection{Distributed Pseudo-tree Optimization Procedure (DPOP)}

\begin{figure}[t]	
\hspace{-2.2cm}
\begin{minipage}[c]{0.4\textwidth}
\centering
	\includegraphics[width=.35\textwidth]{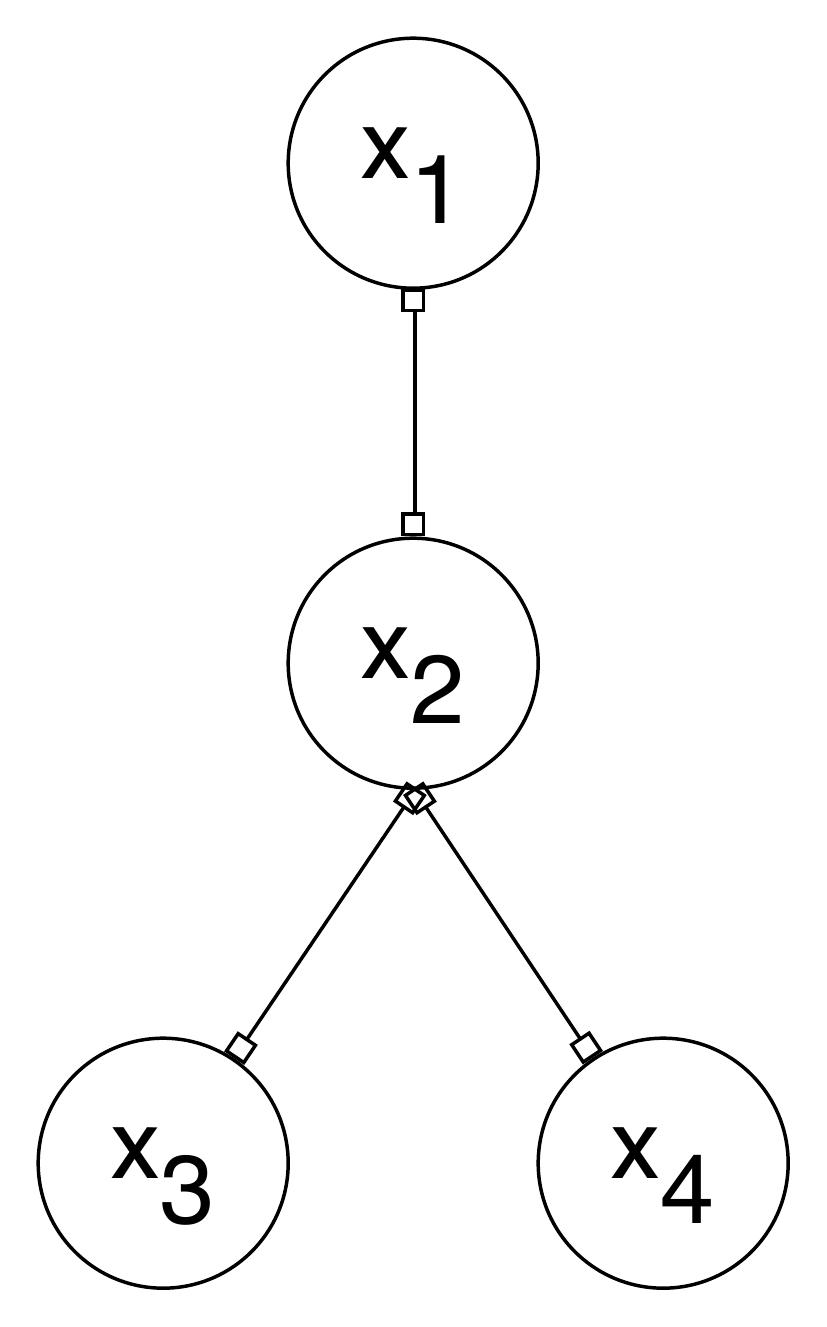}
\end{minipage}	\hspace{-0.8cm}
\begin{minipage}[t]{0.23\textwidth}
		\begin{tabular}{|c|c|c|}
			\hline
			$x_i$ & $x_j$ & $\textit{utility}$\\
			\hline
			0 & 0 & 5\\
			0 & 1 & 8\\
			1 & 0 & 20\\
			1 & 1 & 2\\
			\hline
		\end{tabular}
\end{minipage}	\hspace{.8cm}
\begin{minipage}[c]{0.31\textwidth}

		\begin{tabular}{|c|c|c|}
		\hline
		$x_2$ & $x_1$ & {\it utility} \\
		\hline
		0	& 0	& 5+20+20=45\\
		0	& 1	& 8+20+20=48\\
		1	& 0	& 20+8+8=36\\
		1	& 1 & 2+8+8=18\\
		\hline
		\end{tabular}
\end{minipage}
	\caption{DCOP Example}\label{cg}	
	\end{figure}

\emph{DPOP}~\cite{petcu:05} is one of the most popular complete
algorithms for the distribution resolution of DCOPs; as discussed in several works, it 
has several nice properties (e.g., it requires only a linear number of messages), and it has
been used as the foundations for several more advanced  algorithms~\cite{DBLP:conf/atal/PetcuFP06,DBLP:conf/ijcai/PetcuF07,DBLP:conf/ijcai/PetcuFM07}.

The premise of DPOP is the generation of a \emph{DFS-Pseudo-tree}---composed of a subgraph of 
the constraint graph of a DCOP. The pseudo-tree has a node for each agent in the DCOP;
edges meet the following conditions:
{\bf (a)}
If an edge $(a_1,a_2)$ is present in the pseudo-tree, then there are two variables
	$x_1,x_2$ s.t. $\alpha(x_1) = a_1$, $\alpha(x_2) = a_2$, and
	$(x_1,x_2) \in E_{\cal P}$;
	{\bf (b)} The set of edges describes a rooted tree;
{\bf (c)} For each pair of variables $x_i,x_j$ s.t. $\alpha(x_i) \neq \alpha(x_j)$ and $(x_i,x_j)\in E_{\cal P}$, we have 
that $\alpha(x_i)$ and $\alpha(x_j)$ appear in the same branch of the pseudo-tree. $\alpha(x_i)$ and $\alpha(x_j)$ are also called the pseudo-parent and pseudo-child of each other. 

 Algorithms exist (e.g., \cite{hamadi:98}) to support
the distributed construction of a DFS-Pseudo-tree. 
Given a DCOP $\cal P$, we will refer to a DFS-Pseudo-tree of $\cal P$ by 
${\cal T}_{\cal P} = (A, ET_{\cal P})$. We will
also denote with $a \mapsto_{\cal P} b$ if there exists a sequence of
edges $(a_1,a_2), (a_2,a_3), \dots, (a_{r-1}, a_r)$ in $ET_{\cal P}$ such that $a=a_1$
and $b=a_r$; in this case, we say that $b$ is reachable from $a$ in ${\cal T}_{\cal P}$. Given an
agent $a$, we denote with ${\cal S}_{\cal P}(a)$ the set of agents in 
${\cal T}_{\cal P}$ in the subtree rooted at $a$ (including $a$ itself).

\noindent
The DPOP algorithm operates in two phases:
\begin{list}{$\bullet$}{\itemsep=0pt \parsep=1pt \topsep=1pt} 
\item {\bf UTIL Propagation:} 
During this phase, messages flow bottom-up in the tree, from the 
leaves towards the root. Given a node $N$, the  UTIL message sent by $N$ 
summarizes the maximum utility achievable within the subtree rooted at $N$
for each combination of values of variables belonging to the separator set~\cite{DBLP:books/daglib/0016622} of $N$.
The agent does so by summing the utilities in the UTIL messages received from its 
children agents, and then
projecting out its own variables by optimizing over them.

\item{\bf VALUE Propagation:} 
During this phase, messages flow top-down in the tree. Node
$N$ determines an assignment to its own variables that produces the maximum utility based
on the assignments given by the ancestor nodes; this assignment is then propagated as VALUE
messages to the children.
\end{list} 

Let us consider a DCOP with $X=\{x_1,x_2,x_3,x_4\}$, each with $D_{x_i}=\{0,1\}$ and with binary
	constraints described by the graph (and pseudo-tree) and utility table (assuming $i>j$)
	 in Fig. \ref{cg} (left and middle). For simplicity, we assume a single variable per agent.
	 Node $x_2$ will receive two UTIL
	messages from its children; for example, the message from $x_3$ will indicate that the best utilities
	are $20$ (for $x_2=0$) and $8$ (for $x_2=1$). In turn, $x_2$ will compose the UTIL messages with its own
	constraint, to generate a new utility table, shown 
	in Fig. \ref{cg} (right). This will lead to a UTIL 
	meassage sent to $x_1$  indicating
	utilities of $45$ for $x_1=0$ and $48$ for $x_1=1$. In the VALUE phase, node $x_1$ will generate an 
	assignment of $x_1=1$, which will be sent as a VALUE message to $x_2$; in turn, $x_2$
	 will trigger the assignment $x_2=0$ as a VALUE message to its children.

\vspace{-0.5cm}
\section{Logic-Programing-based DPOP (\clpdcop)}
In this section, we illustrate the \emph{\clpdcop} framework, designed
to map DCOPs into logic programs that can be solved in a distributed manner
using the DPOP algorithm.
 \vskip2ex
\begin{figure}[htbp]
\begin{minipage}[c]{0.45\textwidth}
\includegraphics[width=\textwidth]{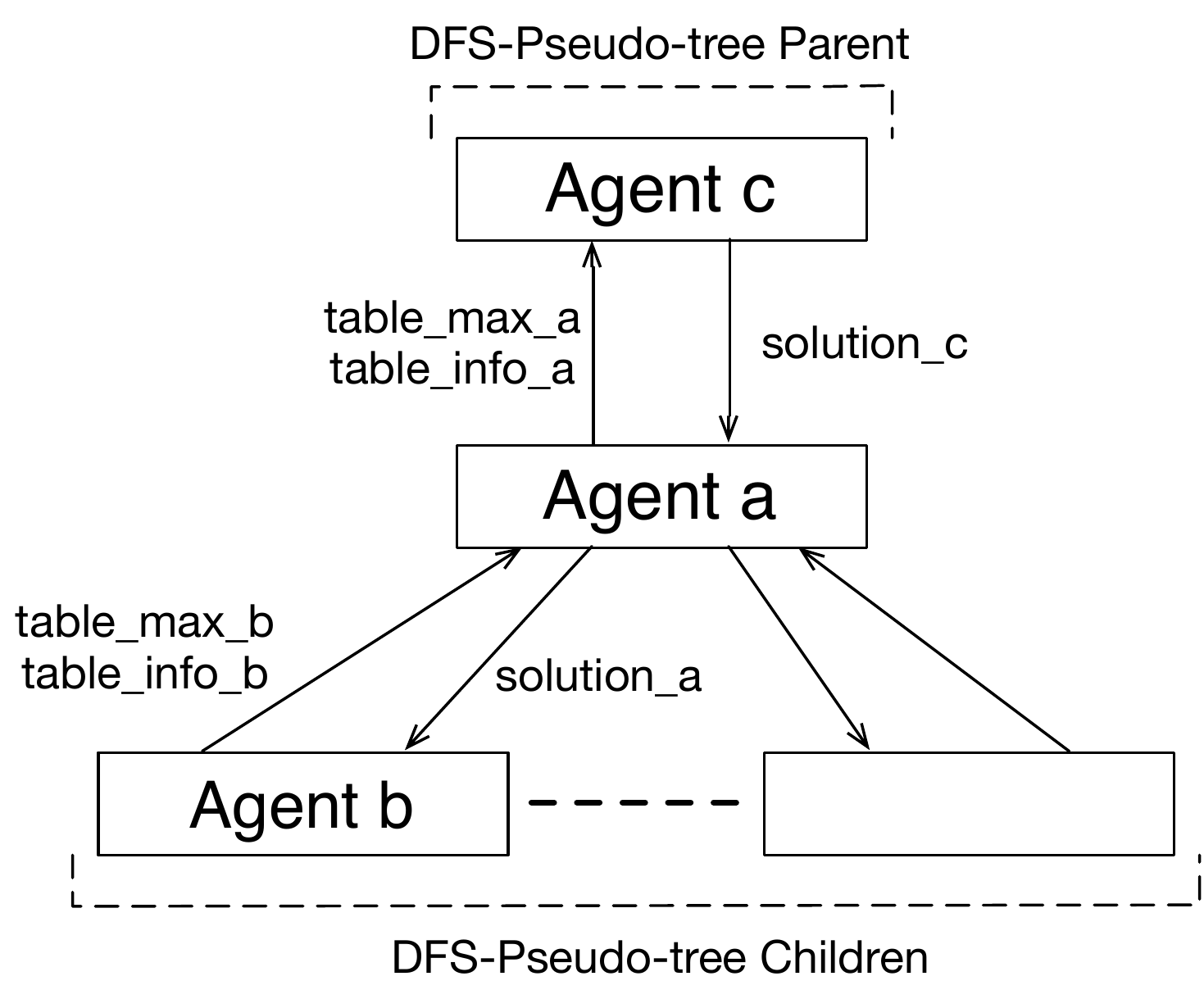}
\caption{Overall Communication Needs}
\label{overall}
\end{minipage}
\begin{minipage}[c]{0.5\textwidth}
\includegraphics[width=\textwidth]{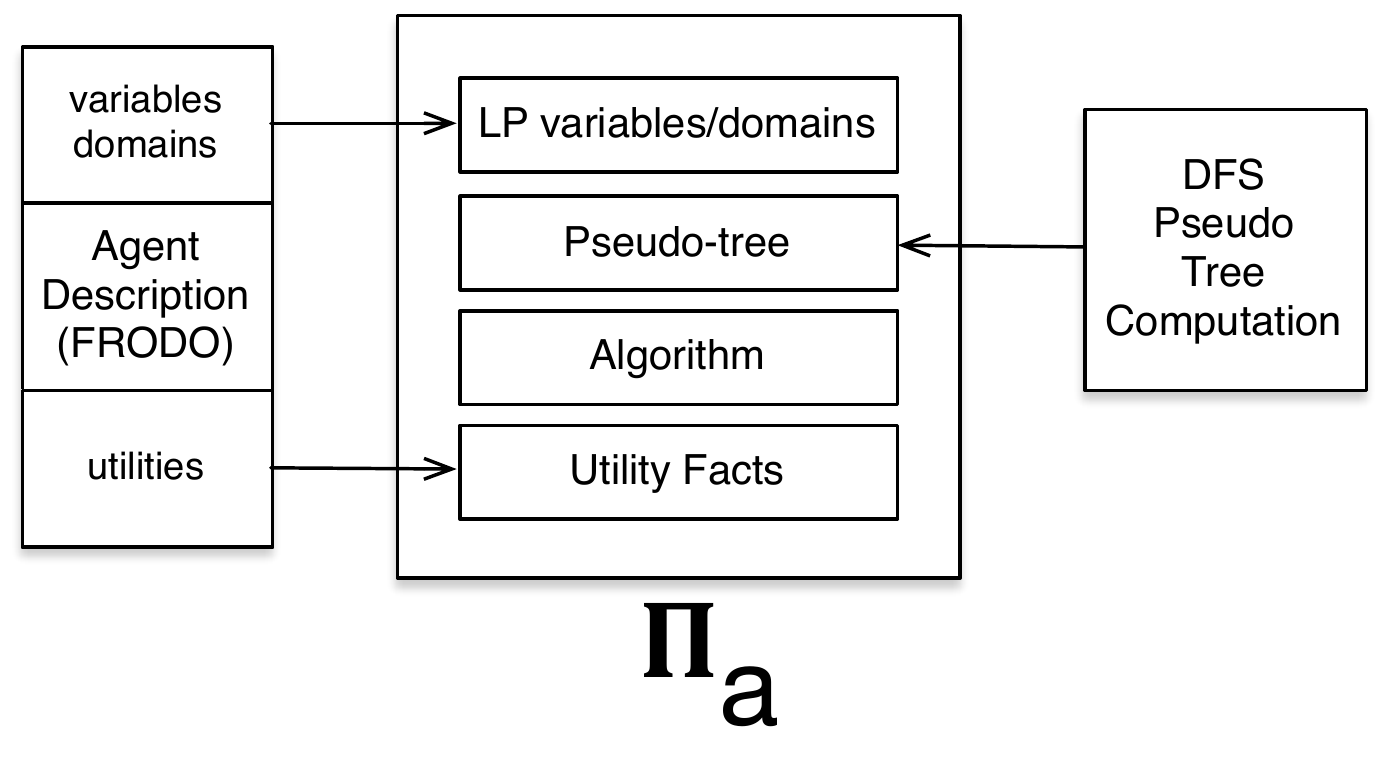}
\caption{Components of an Agent in \clpdcopf}
\label{comm1}
\end{minipage}
\end{figure}

\vspace{-0.4cm}
\subsection{Overall Structure}
The overall structure  of \clpdcop\ is summarized in Fig. \ref{overall}. Intuitively,
each agent $a$ of a DCOP $\cal P$ is mapped to a logic program $\Pi_a$. Agents  exchange information according to the communication protocol of DPOP. These exchanges are represented by collections of facts that are communicated
between agents. In particular,  
\begin{list}{$\bullet$}{\itemsep=0pt \parsep=1pt \topsep=1pt} 
\item UTIL messages from agent $b$ to agent $a$ are encoded as facts 
$\textit{table\_max\_b}( L ),$
where $L$ is a list of $[u, v_1, \dots, v_k]$. Each one is a
row of the UTIL message, where $u$ is the maximum utility for the combination of values $v_1, \dots, v_k$. It is also necessary to transmit an additional message describing
the variables being communicated:
{\small $\textit{table\_info\_b}( [v(x_1,low_1,high_1), \dots, v(x_k,low_k,high_k)] ).$}
This message identifies the names of the variables being communicated and their respective domains. It should be mentioned that the UTIL message from $b$ to $a$ can contain variables belonging to some ancestors of $a$.   

\item VALUE messages from agent $c$ to agent $a$ are encoded as facts 
$\mathit{solution\_c}(\mathit{Var},\mathit{Val}),$
where \emph{Var} is the name of a variable and \emph{Val} is the value assigned to it.
\end{list}

\vspace{-0.4cm}
\subsection{\clpdcop\ Execution Model}

\noindent
{\bf Computing DFS-PseudoTree:}
One can use existing off-the-shelf distributed algorithms to construct pseudo-trees. A commonly used algorithm is the distributed DFS protocol~\cite{hamadi:98}, that  creates a DFS tree with the max-degree heuristic as the variable-ordering heuristic. The max-degree heuristic favors variables with larger numbers of constraints to be higher up in the pseudo-tree.
\smallskip

\noindent{\bf Solving a DCOP:}
The actual agent $a$ is implemented by a logic program $\Pi_a$. In the context
of this paper, the logic program is a CLP program, whose entry point is 
a predicate called {\tt agent}:

{\small
\begin{verbatim}
    agent :- agent(ID),
             (\+is_leaf(ID) -> get_utils; true),
             (\+is_root(ID) -> compute_utils, send_utils, get_value; true),
             (\+is_leaf(ID) -> compute_value, send_value; compute_value).			 
\end{verbatim}}
\noindent
The logic program implements the {\tt compute\_utils} and
the {\tt compute\_value} predicates. They are described in the next section.
\vspace{-0.3cm}
\subsection{Modeling \clpdcop\ as CLP}
In this section, we illustrate the structure of the logic program that encodes
each individual agent. We propose two alternative models. The first one 
follows the model illustrated in Fig~\ref{comm1}: the input DCOP is described
using the standardized format introduced by the FRODO DCOP platform \cite{leaute:09}. 

In the first model, referred to as \clpdcopf, the FRODO model is literally 
translated into collections of logic programming facts. The second model, referred
to as \clpdcopr, follows the more ``realistic'' option of capturing the 
hard constraints present in the DCOP model explicitly as logical constraints, instead
of forcing their mapping to explicit utility tables (as automatically done by FRODO).

\vspace{-0.3cm}
\subsubsection{\clpdcopf}
The logic program $\Pi_a$
modeling an agent is composed of four primary
modules, as illustrated in Fig. \ref{comm1}:
\begin{list}{ }{\topsep=1pt \parsep=0pt \itemsep=1pt}
\item[1.] \emph{Agent, Variables and Domains:} the core components of the agent variables and
	domains are encoded in $\Pi_a$ by facts of the form:
	\begin{list}{$\circ$}{\topsep=1pt \parsep=0pt \itemsep=1pt}
	\item A single fact $agent(a)$ describing the identity of the agent;
	\item For each variable $x_i$ with domain $D_{x_i}$, such that 
	  $\alpha(x_i) = a$ or $\alpha(x_j)=a$ for some variable $x_j$ such that $(x_i,x_j)\in E_{\cal P}$:
	    a fact $variable(x_i,min(D_{x_i}),max(D_{x_i}))$ and 
		a fact $owner(\alpha(x_i), x_i)$.
	\end{list}
\item[2.] \emph{DFS-Pseudo-Tree:} the local position of $a$ in the DFS-Pseudo-tree is described by:
  	  \begin{list}{$-$}{\topsep=1pt \parsep=0pt \itemsep=1pt}
		\item facts of the form $child(b)$ where $b$ is agent s.t. $(a,b)\in ET_{\cal P}$;
		\item a fact $parent(c)$ where $c$ is the (only) agent s.t. $(c,a)\in ET_{\cal P}$; and
		\item a fact $ancestor(c)$ where $c$ is any non-parent ancestor of $a$ in the pseudo-tree, i.e.,
			any agent $c$ s.t. $(c,a) \not\in ET_{\cal P}$ and $c \mapsto_{\cal P} a$.
	  \end{list}
\item[3.] \emph{Utilities/Constraints:} the constraints are obtained as direct translation of 
	the utility tables in the FRODO representation: for each constraint $f_j$, there is 
	a fact of the form $constraint\_f_j(L)$, where $L$ is a list containing lists $[f_j(v_1,\dots v_r),v_1,\dots,v_r]$ 
	for each assignement $\{x_1/v_1, \dots, x_r/v_r\}$ to the variables of
	$scp(f_j) = \{x_1,\dots, x_r\}$ where $f_j(v_1,\dots,v_r) \neq -\infty$.
	Each constraint is further described by the facts: 
{\em (i)} a fact $\textit{constraint}(f_j)$, identifying the name of each constraint, 
{\em (ii)} a fact $\textit{scope}(f_j,x_i)$ for each $x_i \in scp(f_j)$, identifying the variables contributing to the scope of the constraint, and 
{\em (iii)} facts of the form $\textit{constraint\_agent}(f_j,a_r)$, identifying agents that has variables in the scope of the constraint.

\item[4.] \emph{Resolution Engine:} a collection of rules that implement the {\tt compute\_utils} and
	{\tt compute\_value}---these are described below.
\end{list}

The core of the computation of the UTIL message is implemented within the 
{\tt compute\_utils} predicate. Intuitively, the construction of the UTIL message
is mapped to a CLP problem. Its construction and
resolution  can be summarized as follows:
\vspace{-0.2cm}
{\small\begin{verbatim}
   ... define_variables(L,Low,High), 
       define_constraints(L,Util),
       generate_utils(Low,High,UTILITIES), ...
\end{verbatim}}
\vspace{-0.2cm}
The steps can be summarized as follows:
\begin{list}{$\bullet$}{\topsep=1pt \itemsep=1pt \parsep=0pt}
\item The  {\tt define\_variables} predicate is used to collect the variables that belong to
	the agent and its ancestors (returned in the list {\tt Low} and {\tt High}, respectively),
	and for each variable generates a corresponding CLP domain variable. The collecting variables phase is based
	on the {\it variable} facts (describing all variables owned by the agent) and the variables
	indicated in the {\it table\_info\_b} messages received from the children; these may contain variables that belong
	to pseudo-parents in the tree and unknown to the agent $a$. To enable interpretation of the CLP variables,
	two facts {\tt low\_vars}(Low) and {\tt high\_vars}(High) are created in this phase. In the latter phase, for each $X_i$ in the collection of variables collected from the former phase calls $X_i \texttt{ in }\ell..m$
 where $\ell$ and $m$ are the minimum and maximum value of $X_i$'s domain which are either known to the agent or given in received the {\it table\_info\_b} message.
\item The predicate {\tt define\_constraints} creates CLP constraints capturing the utilities the agent has to deal with---these
	include the utilities described by each  $table\_max\_b$ message received from a child $b$ and the utilities $f_j$ of the agent $a$ s.t.
	$scp(f_j)$ does not contain any variables  in $\bigcup_{(a,b)\in ET_{\cal P}} \{x\in X\:|\: \alpha(x)=b\}$.
	For each utility $f_i$ of these utilities (described by a list of lists), the predicate {\tt define\_constraints} introduces a constraint of 
		the form:\\
		\centerline{\tt table([[Ui, X1, .., Xr]] , L, [order(id3), consistency(domain)]) }
	where:
		\begin{list}{$\circ$}{\topsep=0pt \itemsep=0pt \parsep=0pt}
		\item $X1, \dots, Xr$ are the CLP variables which were created by {\tt define\_variables} and correspond to 
			the scope of this utility.
		\item $L$ is the list of lists given in $constraint\_f_i(L)$;
		\item $Ui$ is a new variable introduced for each utility $f_i$.
		\end{list}
	The final step of the {\tt define\_constraints} is to introduce the additional CLP constraint
		$Util \#\!\!= U_1 + U_2 + $\dots$ + U_s$ where $U_i$ are the variables introduced in the {\tt table} constraints
		and $Util$ is a brand new variable.
\item The {\tt generate\_utils} predicate has the following general structure:

{\small\begin{verbatim}
 generate_utils(Lo, Hi, UTILITIES) :-
  findall([Util|Hi], (labeling([],Hi),find_max_util(Lo,Hi,Util)),UTILITIES).
 find_max_util(Lo, Hi, Util) :-
  maximize(labeling([ff],Lo), Util), assert(agent_a_table_max(Lo,Hi)).
\end{verbatim}}
\end{list}

The core of the computation of the VALUE message takes advantage of the fact that the combination of
variables producing the maximum values are asserted as {\tt agent\_a\_table\_max} facts during the UTILs
phase, enabling a simple lookup to compute the solution. This can be summarized as follows:

{\small\begin{verbatim}
   ... high_vars(H),
   findall(Value,(member(Name,H),solution(Name,Value)), Sols),
   agent_a_table_max(Low,Sols),
   low_vars(Lo), length(Lo,Len), I in 1..Len,
   findall(solution(Name,Value), 
           (indomain(I), nth1(I,Lo,Name), nth1(I,Low,Value)), VALUES), ...
\end{verbatim}}
\vspace*{-0.5cm}
\subsubsection{\clpdcopr}
An alternative encoding takes advantage of the fact that the utilities provided in the
utility table of a FRODO encoding are the results of enumerating the solutions of 
\emph{hard constraints}. A hard constraint captures a relation $f_j(x_1,\dots,x_r) \oplus u$ where
$\oplus$ is a relational operator, and $u$ is an integer.
This is typically captured in FRODO as a table, containing all tuples of values
from $D_{x_1}\times\dots\times D_{x_r}$ that satisfy the relation (with a utility value of
$0$), and the default utility value of $-\infty$ assigned to the remaining tuples.

This utility can be directly captured in CLP, thus avoiding the transition through 
the creation of an explicit table of solutions:
\[ hard\_constraint\_f_j(X_1,\dots,X_r) :- \widehat{f_j}(X_1,\dots,X_r) \widehat{\oplus} u \]
where $\widehat{f_j}$ and $\widehat{\oplus}$ are the CLP operators corresponding to 
$f_j$ and $\oplus$.
For example, the smart grid problems used in the experimental section uses  hard constraints
encoded  as \\
\centerline{
	$\mathtt{hard\_constraint\_eq0(X_{1,2}, X_{2,1}) \:\:  {:}{-}  \:\: X_{1,2} + X_{2,1} \#= 0}$
} 
The resulting encoding of the UTIL value computation will modify the encoding of 
\clpdcopf\ as shown below

\begin{center}
\begin{tabular}{lcr}
 \begin{minipage}[c]{.4\textwidth}
	{\small\begin{verbatim}
constraint_f(L), 		
table([[U,X_1,...,X_r]],L,_)
	\end{verbatim}}
\end{minipage} & 
\Large{$\Rightarrow$} &
\begin{minipage}[c]{.4\textwidth}
	{\small \begin{verbatim}
hard_constraint_f(X_1,...,X_r)
	\end{verbatim}}
\end{minipage}
\end{tabular}
\end{center}

\vspace{-0.3cm}
\subsection{Some Implementation Details}
The current implementation of \clpdcop\ makes use
of the Linda \cite{linda} infrastructure 
of SICStus Prolog \cite{sicstus} to handle all the 
communication.

Independent agents can be launched on different machines
and connect to a Linda server started on a dedicated host. Each agent
has a main clause of the type \\
{\tt \small
run\_agent :- prolog\_flag(argv, [Host,Port]),linda\_client(Host:Port),agent.
}

The operations of sending a UTIL message from $b$ to the parent $a$  is simply
realized by a code fragment of the type \\
{\tt \small
send\_util(Vars,Utils,To):- out(msg\_to(To),[table\_info\_b(Vars),table\_max\_b(Utils)]).
} 	
\vskip-0.43cm

The corresponding reception of UTIL message by $a$ will use a predicate of the form \\
{\tt  \small
get\_util(Vars,Utils,Me):- in(msg\_to(Me), [table\_info\_b(Vars),table\_max\_b(Utils)]).
}

The communication of VALUE messages is analogous. 
{\tt get\_value} and {\tt send\_value} are simple wrappers of the predicates discussed above.

\vspace{-0.3cm}
\subsection{Some Theoretical Considerations}
The soundness and completeness of the \clpdcop\ system is a 
natural consequence of the soundness and completeness properties
of the  DPOP algorithm, along with the soundness and completeness of the
CLP(FD) solver of SICStus Prolog. Since \clpdcop\ emulates the computation and communication
operations of DPOP, each $\Pi_a$ program is a correct and
complete implementation of the corresponding agent $a$.

In the worst case, each agent in \clpdcop, like DPOP, needs to compute,
store, and send a utility for each combination of values
of the variables in the separator set of the agent. Therefore,
like DPOP, LP-DPOP also suffers from an exponential
memory requirement, i.e., the memory requirement per agent
is $O(maxDom^w)$, where $maxDom = \argmax_i |D_i|$ and
$w$ is the induced width of the pseudo-tree.
\vspace{-0.4cm}
\section{Experimental Results}
We compare two implementations of the \clpdcop\ framework,
\clpdcopf\ and \clpdcopr\, with a publicly-available implementation of DPOP, which is available on the FRODO framework~\cite{leaute:09}. All experiments are conducted on a Quadcore 3.4GHz machine with 16GB of memory. The runtime of the algorithms are measured using the simulated runtime metric~\cite{sultanik:07}. The timeout is set to 10 minutes.  
Two domains, randomized graphs and smart grids, were used in the experiments.

\smallskip

\noindent
{\bf Randomized Graphs:}
A randomized graph generated using the model in~\cite{erdos59a} with the input parameters $n$ (number of nodes) and $M$ (number of binary edges) will be used as the constraint graph of a DCOP instance $\cal P$.

Each instance ${\cal P}=(X,D,F,A,\alpha)$ is generated using 
five parameters: $|X|$, 
 $|A|$, 
the domain size $d$ of all variables, the constraint density $p_1$ (defined as the ratio between the number of binary edges $M$ and the maximum number of binary edges among $|X|$ nodes), 
and the constraint tightness $p_2$ (defined as the ratio between the number of infeasible value combinations, that is, their utility equals $-\infty$, and the total number of value combinations).

We conduct experiments, where we vary one parameter in each experiment. The ``default'' value for each experiment is $|A| = 5$, $|X| = 15$, $d=6$, $p_1=0.6$, and $p_2=0.6$. As the utility tables of instances of this domain are randomly generated, the programs for \clpdcopr\ and \clpdcopf\ are very similar. Thus, we only compare FRODO with \clpdcopf. Table~\ref{randomGraphResults} shows the percentage of instances solved and the average simulated runtime (in ms) for the solved instances; each data point is an average over 50 randomly generated instances. If an algorithm fails to solve more than $85\%$ of instances in a specific configuration, then we consider that it fails to solve problems with that configuration. 

\renewcommand{\arraystretch}{0.5}
\begin{table}
	\caption{Experimental Results on Random Graphs (\%: Solved; Time: Runtime)}
	\label{randomGraphResults}
	\begin{minipage}{\textwidth}
		\begin{tabular}{|c|r|r|r|r||c|r|r|r|r|}
			\hline\hline
 \multirow{2}{*}{$| {X}|$}   	& \multicolumn{2}{c|}{DPOP} & \multicolumn{2}{c||}{\clpdcopf}&
 \multirow{2}{*}{$d$}   	& \multicolumn{2}{c|}{DPOP} & \multicolumn{2}{c|}{\clpdcopf} \\
	& \%  & Time & \%    & Time  & 
	& \%   & Time & \%    & Time \\
\hline
     5    &    100\%	&    35 	 & 100\%    & 30 & 
           4    &    100\%     &    782  & 100\%    & 74\\ 
     10  &    100\%	&    204 	 & 100\%    & 264 & 
            6    &     90\%      &    28,363  & 100\%   & 539\\  
     15    &  86\% 	&    39,701 	 & 100\%    & 1,008 & 
             8    &     14\%      &    -  & 98\%    & 22,441\\
     20    &   0\%    	&    - 		 & 100\%    & 1,263 & 
             10   &     0\%       &    -  & 94\%    & 85,017\\  
     25    &     0\%     &    - 		 & 100\%    & 723 & 
             12   &     0\%       &    -  & 60\%    & -\\ 
     30    &     0\%     &    -  		&  100\%    & 255 & 
          & & & & \\
     35    &     0\%      &    -  	&  100\%    & 256 & 
          & & & & \\
          \hline \hline
  \multirow{2}{*}{$p_1$} & \multicolumn{2}{c|}{DPOP} & \multicolumn{2}{c||}{\clpdcopf}   & 
  \multirow{2}{*}{$p_2$} & \multicolumn{2}{c|}{DPOP} & \multicolumn{2}{c|}{\clpdcopf} \\
	& \%  & Time & \%    & Time  & 
	& \%   & Time & \%    & Time \\
     \hline
   0.3    &    100\%     &    286  & 100\%    & 2,629 & 
           0.4    &   86\%      &    48,632  & 92\%    & 155,089\\ 
     0.4    &     100\%     &    1,856  & 100\%    & 2,038 & 
              0.5    &    94\%      &    38,043  & 100\%    & 23,219\\ 
     0.5    &     100\%       &    13,519  & 100\%    & 938 & 
               0.6    &     90\%      &    31,513  & 100\%    & 844\\ 	 
     0.6    &     94\%       &    42,010  & 100\%    & 706 & 
              0.7    &  90\% 	&    39,352  & 100\%    & 84\\ 
     0.7    &     56\%       &   -  & 100\%    & 203 & 
             0.8    &    92\%       &    40,525  & 100\%    & 61\\ 
     0.8    &     20\%      &    -  & 100\%   & 176 & 

            0.9    &     96\%       &    27,416  & 100\%   & 60\\  	
\hline\hline
		\end{tabular}
	\end{minipage}
\end{table}

The results show that \clpdcopf\ is able to solve more problems and is faster than DPOP when the problem becomes more complex (i.e., increasing ${|X|}$, $d$, $p_1$, or $p_2$). The reason is that at a specific percentage of hard constraints (i.e., $p_2 = 0.6$), \clpdcopf\ is able to prune a significant portion of the search space. Unlike DPOP, \clpdcopf\ does not need to explicitly represent the rows in the UTIL table that are infeasible, resulting in lower memory usage  and runtime needed to search through search space. The size of the search space pruned increases as the complexity of the instance grows, making the difference between the runtimes of \clpdcopf\ and DPOP  significant. 

\smallskip
\noindent
{\bf Smart Grids:}
A {\em customer-driven microgrid} (CDMG), one possible instantiation of the smart grid problem, has recently been shown to subsume several classical power system sub-problems (e.g., load shedding, demand response, restoration)~\cite{jain:12}.
In this domain, each agent represents a node with 
consumption, generation, and transmission preference, and a global cost function. Constraints include the power balance and 
no power loss  principles, the generation and consumption limits,  
and the capacity of the power line between nodes. The objective 
is to minimize a global cost function. CDMG optimization problems are well-suited to be modeled with DCOPs due to their distributed nature. Moreover, as some of the constraints in CDMGs (e.g., the power balance principle) can be described in functional form, they can be exploited by \clpdcopr. For this reason, both \clpdcopf\ and \clpdcopr\ were used in this domain.

\begin{figure}
    \parbox{2.5in}{
        \centering \footnotesize
        \includegraphics[width=2.0in]{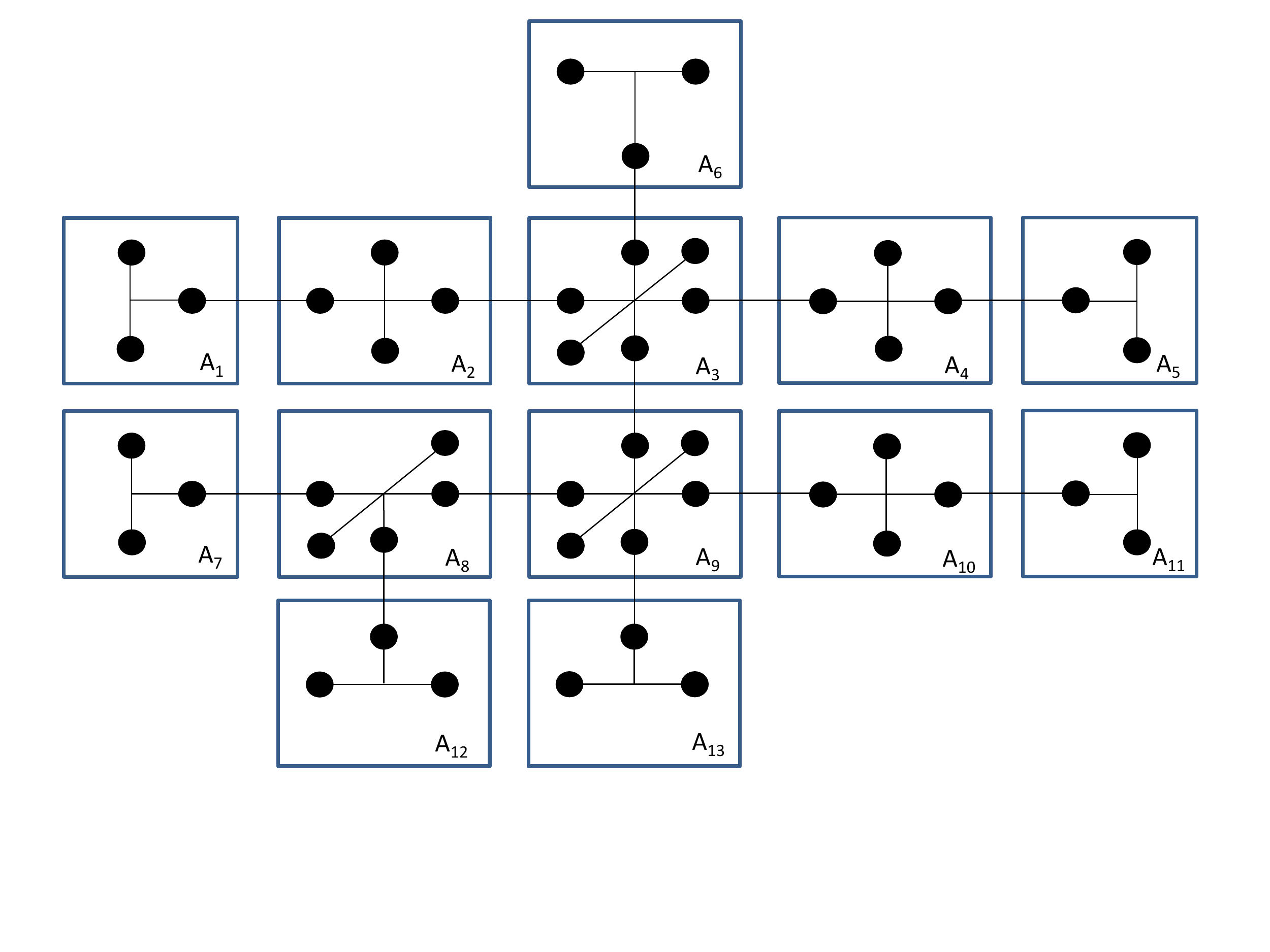}
        \\ \vspace{-2.0em} \hspace{1em} (a) IEEE Standard 13 Bus Topology} \hspace{-0em}
    \parbox{1.5in}{
        \centering \footnotesize
        \includegraphics[width=1.9in]{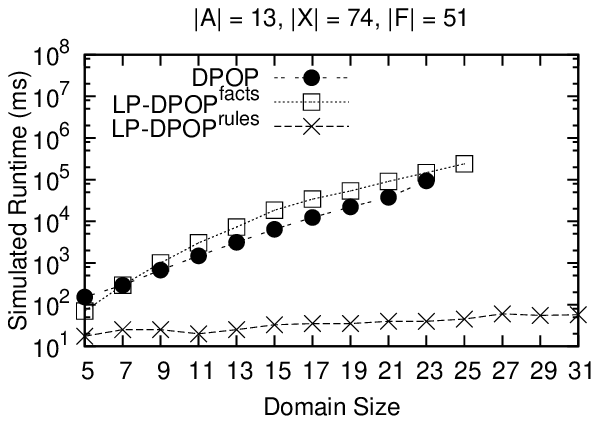}
        \\ \vspace{-0.5em} \hspace{1em} (b) 13 Bus Topology} \hspace{-0em}
    \parbox{1.5in}{
        \centering \footnotesize
        \includegraphics[width=1.9in]{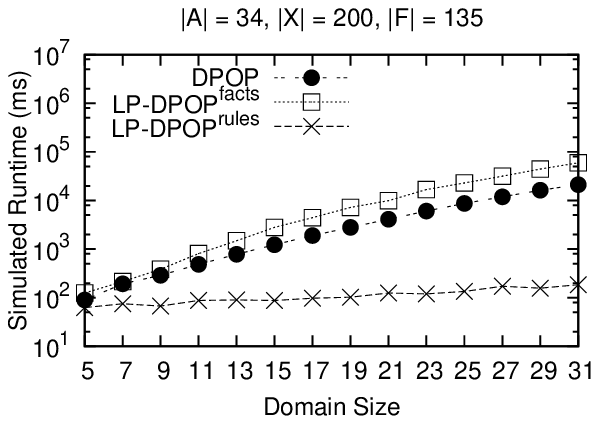}
        \\ \vspace{-0.5em} \hspace{1em} (c) 34 Bus Topology} \hspace{1.5em}
    \parbox{1.5in}{
        \centering \footnotesize
        \includegraphics[width=1.9in]{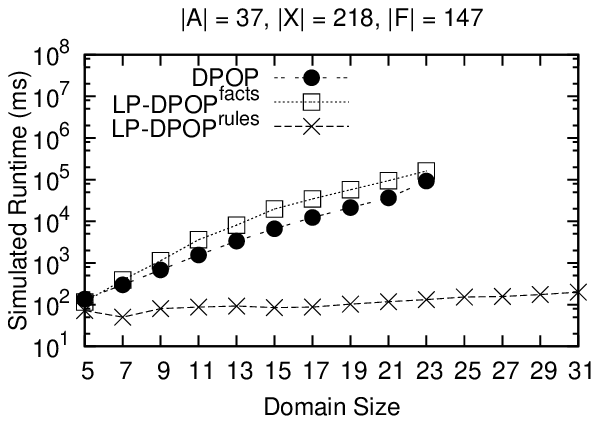}
        \\ \vspace{-0.5em} \hspace{1em} (d) 37 Bus Topology} \hspace{1.5em}
\parbox{1.5in}{
        \centering \footnotesize
        \includegraphics[width=1.9in]{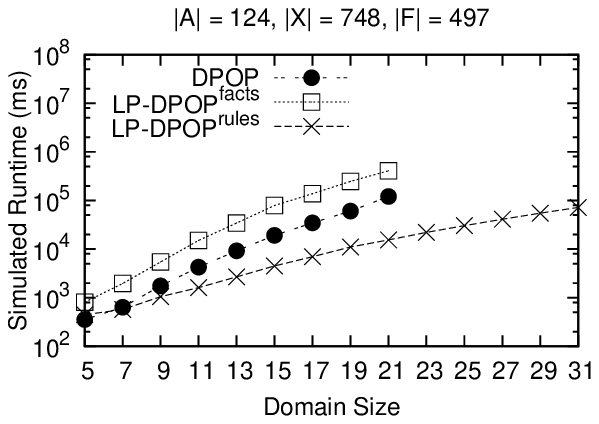}
        \\ \vspace{-0.5em} \hspace{1em} (e) 123 Bus Topology}
\vspace{-0.25em}
\caption{Experiment Results on Smart Grids}
\label{resultSmartgrid}
\end{figure}

 We conduct experiments on a range of CDMG problem instances generated using the four network topologies following the IEEE standards and varying the domain of the variables.\footnote{\url{www.ewh.ieee.org/soc/pes/dsacom/}}  
Fig.~\ref{resultSmartgrid}(a) displays the topology of the IEEE 13 Bus network, where rectangles represent nodes/agents, filled circles represent variables, and links between variables represent constraints. The initial configuration of the CDMG and the precise equations used in the generation of the problems can be found in \cite{jain:12}. 
The experimental results for the four largest standards, the 13, 34, 37, and 123 Bus Topology,\footnote{In 123 Bus Topology's experiments, a multi-server version of \clpdcopf\ and \clpdcopr\ was used because of the limit on the number of concurrent streams supported by Linda and SICStus. FRODO cannot be run on multiple machines.} are shown in Fig.~\ref{resultSmartgrid}(b),~\ref{resultSmartgrid}(c),~\ref{resultSmartgrid}(d), and~\ref{resultSmartgrid}(e), respectively. We make the following observations:   
\begin{list}{$\bullet$}{\itemsep=0pt \parsep=1pt \topsep=1pt} 
\item \clpdcopr\ is the best among the three systems both in terms of runtime and scalability in all experiments. 
\clpdcopr 's memory requirement during its execution is significant smaller and increases at a much slower pace than other systems. This indicates that the rules used in expressing the constraints help the constraint solver to more effectively prune the search space resulting in a better performance.  
\item \clpdcopf\ is slower than DPOP in all experiments in this domain. It is because \clpdcopf\ often needs to backtrack while computing the UTIL message, and each backtracking step requires the look up of several related utility tables---some tables can contain many tuples (e.g., one agent in the 13 Bus problem with domain size of 23 could have $3,543,173$ facts). We believe that this is the source of the weak performance of \clpdcopf. 
\end{list}
 
\vspace{-0.3cm}
\section{Conclusion and Future Work}
In this paper, we presented a generic infrastructure built on logic programming to address problems in the 
area of DCOP. The use of a generic CLP solver to implement the individual agents proved to 
be a winning option, largely outperforming existing DCOP technology in terms of speed and scalability. The paper also makes
the preliminary case for a different encoding of DCOPs w.r.t. existing technologies; the ability to explicitly model hard
constraints provides agents with additional knowledge that can be used to prune the search space, further enhancing performance.

This is, in many regards, a preliminary effort that will be expanded in several directions. First, we believe that
different types of DCOP problems may benefit from different types of local solvers within each agent; we currently explore
the use of ASP as an alternative for the encoding the agents.
The preliminary results are competitive and superior to those produced by
DPOP. Classifying DCOP problems in such a way to enable the automated selection of what type of LP-based solver to use is an
open research question to be addressed. The strong results observed in the use of implicit encodings of hard constraints also suggest
the need of developing \emph{DCOP description languages} that separate hard and soft constraints and do not require the explicit
representation for all constraints. 

On the other direction, we view this work as a feasibility study towards the development of distributed LP models (e.g., Distributed ASP). Paradigms like ASP are highly suitable to capture the description of individual agents operating in multi-agent environments; yet, ASP does not inherently provide the capability of handling a distributed ASP computation with properties analogous to those found in DCOP. We believe the models and infrastructure described in this paper could represent the first step in the direction of creating the foundations of DASP and other distributed logic programming models.

\newpage
\bibliographystyle{acmtrans}
\bibliography{../../bibtex/bib2010,../../bibtex/bibfile}

\begin{thebibliography}{}

\bibitem[\protect\citeauthoryear{{Carlsson et al.}}{{Carlsson et
  al.}}{2012}]{sicstus}
{\sc {Carlsson et al.}, M.} 2012.
\newblock {SICStus Prolog User's Manual}.
\newblock Tech. rep., Swedish Institute of Computer Science.

\bibitem[\protect\citeauthoryear{Carriero, Gelernter, Mattson, and
  Sherman}{Carriero et~al\mbox{.}}{1994}]{linda}
{\sc Carriero, N.}, {\sc Gelernter, D.}, {\sc Mattson, T.}, {\sc and} {\sc
  Sherman, A.} 1994.
\newblock {The Linda Alternative to Message Passing Systems}.
\newblock {\em Parallel Computing\/}~{\em 20,\/}~4, 633--655.

\bibitem[\protect\citeauthoryear{Dechter}{Dechter}{2003}]{DBLP:books/daglib/0016622}
{\sc Dechter, R.} 2003.
\newblock {\em Constraint processing}.
\newblock Elsevier Morgan Kaufmann.

\bibitem[\protect\citeauthoryear{Erd\"{o}s and R\'{e}nyi}{Erd\"{o}s and
  R\'{e}nyi}{1959}]{erdos59a}
{\sc Erd\"{o}s, P.} {\sc and} {\sc R\'{e}nyi, A.} 1959.
\newblock On random graphs I.
\newblock {\em Publicationes Mathematicae Debrecen\/}~{\em 6}, 290.

\bibitem[\protect\citeauthoryear{Ezzahir, Bessiere, Belaissaoui, and
  Bouyakhf}{Ezzahir et~al\mbox{.}}{2007}]{ezzahir:07}
{\sc Ezzahir, R.}, {\sc Bessiere, C.}, {\sc Belaissaoui, M.}, {\sc and} {\sc
  Bouyakhf, E.~H.} 2007.
\newblock {DisChoco}: A platform for distributed constraint programming.
\newblock In {\em Proceedings of the Distributed Constraint Reasoning
  Workshop}. 16--27.

\bibitem[\protect\citeauthoryear{Farinelli, Rogers, Petcu, and
  Jennings}{Farinelli et~al\mbox{.}}{2008}]{farinelli:08}
{\sc Farinelli, A.}, {\sc Rogers, A.}, {\sc Petcu, A.}, {\sc and} {\sc
  Jennings, N.} 2008.
\newblock Decentralised coordination of low-power embedded devices using the
  {Max-Sum} algorithm.
\newblock In {\em Proceedings of the International Joint Conference on
  Autonomous Agents and Multiagent Systems (AAMAS)}. 639--646.

\bibitem[\protect\citeauthoryear{Fitzpatrick and Meertens}{Fitzpatrick and
  Meertens}{2003}]{fitzpatrick:03}
{\sc Fitzpatrick, S.} {\sc and} {\sc Meertens, L.} 2003.
\newblock Distributed coordination through anarchic optimization.
\newblock In {\em Distributed Sensor Networks: A Multiagent Perspective},
  {V.~Lesser}, {C.~Ortiz}, {and} {M.~Tambe}, Eds. Kluwer, 257--295.

\bibitem[\protect\citeauthoryear{Gershman, Meisels, and Zivan}{Gershman
  et~al\mbox{.}}{2009}]{gershman:09}
{\sc Gershman, A.}, {\sc Meisels, A.}, {\sc and} {\sc Zivan, R.} 2009.
\newblock {Asynchronous Forward-Bounding} for distributed {COP}s.
\newblock {\em Journal of Artificial Intelligence Research\/}~{\em 34}, 61--88.

\bibitem[\protect\citeauthoryear{Gupta, Jain, Yeoh, Ranade, and Pontelli}{Gupta
  et~al\mbox{.}}{2013}]{gupta:13}
{\sc Gupta, S.}, {\sc Jain, P.}, {\sc Yeoh, W.}, {\sc Ranade, S.}, {\sc and}
  {\sc Pontelli, E.} 2013.
\newblock Solving customer-driven microgrid optimization problems as {DCOP}s.
\newblock In {\em Proceedings of the Distributed Constraint Reasoning
  Workshop}. 45--59.

\bibitem[\protect\citeauthoryear{Hamadi, Bessi\`{e}re, and Quinqueton}{Hamadi
  et~al\mbox{.}}{1998}]{hamadi:98}
{\sc Hamadi, Y.}, {\sc Bessi\`{e}re, C.}, {\sc and} {\sc Quinqueton, J.} 1998.
\newblock Distributed intelligent backtracking.
\newblock In {\em Proceedings of the European Conference on Artificial
  Intelligence (ECAI)}. 219--223.

\bibitem[\protect\citeauthoryear{Jain and Ranade}{Jain and
  Ranade}{2009}]{jain:09}
{\sc Jain, P.} {\sc and} {\sc Ranade, S.} 2009.
\newblock Capacity discovery in customer-driven micro-grids.
\newblock In {\em Proceedings of the North American Power Symposium (NAPS)}.
  1--6.

\bibitem[\protect\citeauthoryear{Jain, Ranade, Gupta, and Pontelli}{Jain
  et~al\mbox{.}}{2012}]{jain:12}
{\sc Jain, P.}, {\sc Ranade, S.}, {\sc Gupta, S.}, {\sc and} {\sc Pontelli, E.}
  2012.
\newblock Optimum operation of a customer-driven microgrid: A comprehensive
  approach.
\newblock In {\em Proceedings of International Conference on Power Electronics, Drives and Energy Systems (PEDES)}. 2012. 1--6.

\bibitem[\protect\citeauthoryear{Kiekintveld, Yin, Kumar, and
  Tambe}{Kiekintveld et~al\mbox{.}}{2010}]{kiekintveld:10}
{\sc Kiekintveld, C.}, {\sc Yin, Z.}, {\sc Kumar, A.}, {\sc and} {\sc Tambe,
  M.} 2010.
\newblock Asynchronous algorithms for approximate distributed constraint
  optimization with quality bounds.
\newblock In {\em Proceedings of the International Joint Conference on
  Autonomous Agents and Multiagent Systems (AAMAS)}. 133--140.

\bibitem[\protect\citeauthoryear{Kumar, Faltings, and Petcu}{Kumar
  et~al\mbox{.}}{2009}]{kumar:09}
{\sc Kumar, A.}, {\sc Faltings, B.}, {\sc and} {\sc Petcu, A.} 2009.
\newblock Distributed constraint optimization with structured resource
  constraints.
\newblock In {\em Proceedings of the International Joint Conference on
  Autonomous Agents and Multiagent Systems (AAMAS)}. 923--930.

\bibitem[\protect\citeauthoryear{L{\'e}aut{\'e} and Faltings}{L{\'e}aut{\'e}
  and Faltings}{2011}]{leaute:11}
{\sc L{\'e}aut{\'e}, T.} {\sc and} {\sc Faltings, B.} 2011.
\newblock Coordinating logistics operations with privacy guarantees.
\newblock In {\em Proceedings of the International Joint Conference on
  Artificial Intelligence (IJCAI)}. 2482--2487.

\bibitem[\protect\citeauthoryear{L{\'e}aut{\'e}, Ottens, and
  Szymanek}{L{\'e}aut{\'e} et~al\mbox{.}}{2009}]{leaute:09}
{\sc L{\'e}aut{\'e}, T.}, {\sc Ottens, B.}, {\sc and} {\sc Szymanek, R.} 2009.
\newblock {FRODO} 2.0: An open-source framework for distributed constraint
  optimization.
\newblock In {\em Proceedings of the Distributed Constraint Reasoning
  Workshop}. 160--164.

\bibitem[\protect\citeauthoryear{Maheswaran, Tambe, Bowring, Pearce, and
  Varakantham}{Maheswaran et~al\mbox{.}}{2004}]{maheswaran:04a}
{\sc Maheswaran, R.}, {\sc Tambe, M.}, {\sc Bowring, E.}, {\sc Pearce, J.},
  {\sc and} {\sc Varakantham, P.} 2004.
\newblock Taking {DCOP} to the real world: Efficient complete solutions for
  distributed event scheduling.
\newblock In {\em Proceedings of the International Joint Conference on
  Autonomous Agents and Multiagent Systems (AAMAS)}. 310--317.

\bibitem[\protect\citeauthoryear{Modi, Shen, Tambe, and Yokoo}{Modi
  et~al\mbox{.}}{2005}]{modi:05}
{\sc Modi, P.}, {\sc Shen, W.-M.}, {\sc Tambe, M.}, {\sc and} {\sc Yokoo, M.}
  2005.
\newblock {ADOPT}: Asynchronous distributed constraint optimization with
  quality guarantees.
\newblock {\em Artificial Intelligence\/}~{\em 161,\/}~1--2, 149--180.

\bibitem[\protect\citeauthoryear{Nguyen, Yeoh, and Lau}{Nguyen
  et~al\mbox{.}}{2013}]{nguyen:13}
{\sc Nguyen, D.~T.}, {\sc Yeoh, W.}, {\sc and} {\sc Lau, H.~C.} 2013.
\newblock Distributed {G}ibbs: A memory-bounded sampling-based {DCOP}
  algorithm.
\newblock In {\em Proceedings of the International Joint Conference on
  Autonomous Agents and Multiagent Systems (AAMAS)}. 167--174.

\bibitem[\protect\citeauthoryear{Ottens, Dimitrakakis, and Faltings}{Ottens
  et~al\mbox{.}}{2012}]{ottens:12}
{\sc Ottens, B.}, {\sc Dimitrakakis, C.}, {\sc and} {\sc Faltings, B.} 2012.
\newblock {DUCT}: An upper confidence bound approach to distributed constraint
  optimization problems.
\newblock In {\em Proceedings of the AAAI Conference on Artificial Intelligence
  (AAAI)}. 528--534.

\bibitem[\protect\citeauthoryear{Petcu and Faltings}{Petcu and
  Faltings}{2005}]{petcu:05}
{\sc Petcu, A.} {\sc and} {\sc Faltings, B.} 2005.
\newblock A scalable method for multiagent constraint optimization.
\newblock In {\em Proceedings of the International Joint Conference on
  Artificial Intelligence (IJCAI)}. 1413--1420.

\bibitem[\protect\citeauthoryear{Petcu and Faltings}{Petcu and
  Faltings}{2007}]{DBLP:conf/ijcai/PetcuF07}
{\sc Petcu, A.} {\sc and} {\sc Faltings, B.} 2007.
\newblock MB-DPOP: A new memory-bounded algorithm for distributed optimization.
\newblock In {\em Proceedings of the International Joint Conference on
  Artificial Intelligence (IJCAI)}. 1452--1457.

\bibitem[\protect\citeauthoryear{Petcu, Faltings, and Mailler}{Petcu
  et~al\mbox{.}}{2007}]{DBLP:conf/ijcai/PetcuFM07}
{\sc Petcu, A.}, {\sc Faltings, B.}, {\sc and} {\sc Mailler, R.} 2007.
\newblock PC-DPOP: A new partial centralization algorithm for distributed
  optimization.
\newblock In {\em Proceedings of the International Joint Conference on
  Artificial Intelligence (IJCAI)}. 167--172.

\bibitem[\protect\citeauthoryear{Petcu, Faltings, and Parkes}{Petcu
  et~al\mbox{.}}{2006}]{DBLP:conf/atal/PetcuFP06}
{\sc Petcu, A.}, {\sc Faltings, B.}, {\sc and} {\sc Parkes, D.~C.} 2006.
\newblock MDPOP: Faithful distributed implementation of efficient social choice
  problems.
\newblock In {\em Proceedings of the International Joint Conference on
  Autonomous Agents and Multiagent Systems (AAMAS)}. 1397--1404.

\bibitem[\protect\citeauthoryear{Prosser}{Prosser}{1996}]{Prosser96}
{\sc Prosser, P.} 1996.
\newblock An empirical study of phase transitions in binary constraint
  satisfaction problems.
\newblock {\em Artificial Intelligence\/}~{\em 81,\/}~1-2, 81--109.

\bibitem[\protect\citeauthoryear{Stranders, Farinelli, Rogers, and
  Jennings}{Stranders et~al\mbox{.}}{2009}]{stranders:09b}
{\sc Stranders, R.}, {\sc Farinelli, A.}, {\sc Rogers, A.}, {\sc and} {\sc
  Jennings, N.} 2009.
\newblock Decentralised coordination of mobile sensors using the {Max-Sum}
  algorithm.
\newblock In {\em Proceedings of the International Joint Conference on
  Artificial Intelligence (IJCAI)}. 299--304.

\bibitem[\protect\citeauthoryear{Sultanik, Lass, and Regli}{Sultanik
  et~al\mbox{.}}{2007}]{sultanik:07}
{\sc Sultanik, E.}, {\sc Lass, R.}, {\sc and} {\sc Regli, W.} 2007.
\newblock {DCOPolis}: a framework for simulating and deploying distributed
  constraint reasoning algorithms.
\newblock In {\em Proceedings of the Distributed Constraint Reasoning
  Workshop}.

 \bibitem[\protect\citeauthoryear{Vinyals, Rodríguez-Aguilar, and
  Cerquides}{Vinyals et~al\mbox{.}}{2009}]{conf/atal/VinyalsRC09}
{\sc Vinyals, M.}, {\sc Rodríguez-Aguilar, J.~A.}, {\sc and} {\sc Cerquides,
  J.} 2009.
\newblock Generalizing DPOP: Action-GDL, a new complete algorithm for DCOPs.
\newblock In {\em Proceedings of the International Joint Conference on
  Autonomous Agents and Multiagent Systems (AAMAS)}. 1239--1240.
  
 \bibitem[\protect\citeauthoryear{Yeoh, Felner, and Koenig}{Yeoh
  et~al\mbox{.}}{2010}]{DBLP:journals/jair/YeohFK10}
{\sc Yeoh, W.}, {\sc Felner, A.}, {\sc and} {\sc Koenig, S.} 2010.
\newblock BnB-ADOPT: An asynchronous branch-and-bound DCOP algorithm.
\newblock {\em Journal of Artificial Intelligence Research\/}~{\em 38},
  85--133.

\bibitem[\protect\citeauthoryear{Yeoh and Yokoo}{Yeoh and
  Yokoo}{2012}]{yeoh:12}
{\sc Yeoh, W.} {\sc and} {\sc Yokoo, M.} 2012.
\newblock Distributed problem solving.
\newblock {\em AI Magazine\/}~{\em 33,\/}~3, 53--65.


\bibitem[\protect\citeauthoryear{Zhang, Wang, Xing, and Wittenberg}{Zhang
  et~al\mbox{.}}{2005}]{zhang:05}
{\sc Zhang, W.}, {\sc Wang, G.}, {\sc Xing, Z.}, {\sc and} {\sc Wittenberg, L.}
  2005.
\newblock Distributed stochastic search and distributed breakout: Properties,
  comparison and applications to constraint optimization problems in sensor
  networks.
\newblock {\em Artificial Intelligence\/}~{\em 161,\/}~1--2, 55--87.

\bibitem[\protect\citeauthoryear{Zivan, Yedidsion, Okamoto, Glinton, and Sycara}{Zivan
  et~al\mbox{.}}{2009}]{zivan:09}
{\sc Zivan, R.}, {\sc Yedidsion, H.}, {\sc Okamoto, S.}, {\sc Glinton, R.}, {\sc and} {\sc Sycara, K.} 2014.
\newblock Distributed constraint optimization for teams of mobile sensing
  agents.
\newblock {\em Autonomous Agents and Multi-Agent Systems\/}, 1--42.


\end{thebibliography}

\label{lastpage}
\end{document}